\documentclass[]{aa}
\usepackage{graphicx}
\usepackage{natbib}

\def\hii{H~{\sc II}}

\def\hh{H$_2$}

\def\hho{H$_2$O}

\def\fsec{\mbox{$.\!\!^{\mathrm s}$}}
\def\farcs{\mbox{$.\!\!^{\prime\prime}$}}
\def\fdeg{\mbox{$.\!\!^{\circ}$}}
\def\gtsim{{_>\atop{^\sim}}}
\def\ltsim{{_<\atop{^\sim}}}

\def\kms{km~s$^{-1}$}

\def\ccm{cm$^{-3}$}

\def\mic{$\mu$m}

\def\msol{M$_{\odot}$}
\def\lsol{L$_{\odot}$}

\def\aap{{A\&A}}

\def\apjl{{ApJ}}
\def\apjs{{ApJS}}

\renewcommand{\citep}[1]{(\citeauthor{#1} \citeyear{#1})}

\begin{document}

\title{Subarcsecond mid-infrared and radio observations of the W3~IRS5 protocluster} 

\author{F. F. S. van der Tak\inst{1} \and P. G. Tuthill\inst{2} \and W. C. Danchi\inst{3} }

\institute{
  Max-Planck-Institut f\"ur Radioastronomie, Auf dem H\"ugel 69, 53121 Bonn, Germany 
  \and School of Physics, University of Sydney, NSW 2006, Australia 
  \and NASA Goddard Space Flight Center, Infrared Astrophysics, Code 685, Greenbelt MD 20771, USA}

\titlerunning{Subarcsecond imaging of W3 IRS5}
\authorrunning{van der Tak et al.}
\offprints{F. F. S. van der Tak, \\ \email{vdtak@mpifr-bonn.mpg.de}}

\date{Received 5 July 2004 / Accepted 27 October 2004}

\abstract{ Observations at mid-infrared (4.8--17.65~\mic) and radio
  (0.7--1.3~cm) wavelengths are used to constrain the structure of the
  high-mass star-forming region W3 IRS5 on 0\farcs1 (200~AU) scales.  Two
  bright mid-infrared sources are detected, as well as diffuse emission. The
  bright sources have associated compact radio emission and probably are young
  high-mass stars.  
  The measured sizes and estimated temperatures indicate that these
  sources together can supply the observed far-infrared luminosity.
  However, an optically thick radio source with a possible
  mid-infrared counterpart may also contribute significant luminosity;
  if so, it must be extremely deeply embedded.
  The infrared colour temperatures of 350--390~K and low radio
  brightness suggest gravitational confinement of the \hii\ regions and
  ongoing accretion at a rate of a few 10$^{-8}$ \msol~yr$^{-1}$ or more.
  Variations in the accretion rate would explain the observed radio
  variability.
  The low estimated foreground extinction suggests the existence of a
  cavity around the central stars, perhaps blown by stellar winds.
  At least three radio sources without mid-infrared counterparts
  appear to show proper motions of $\sim$100~\kms, and may be deeply
  embedded young runaway OB stars, but more likely are clumps in the
  ambient material which are shock-ionized by the OB star winds.


}
  
\maketitle

\section{Introduction}
\label{s:intro}

\begin{figure*}[htb]
\includegraphics[width=8cm,angle=-90]{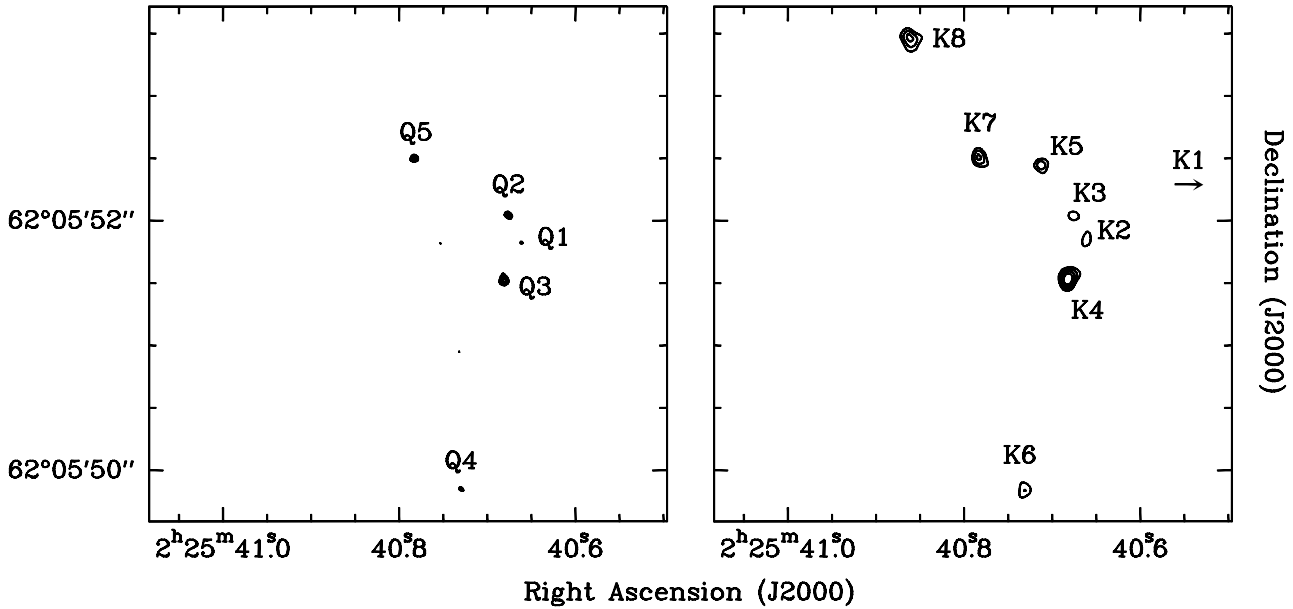}
\caption{Images of W3 IRS5 at 43~GHz (left) and 22~GHz
  (right). Contours start at 4$\sigma$ and increase by 2$\sigma$,
  where $\sigma$=0.16 mJy/beam at 43~GHz and 0.15 mJy/beam at 22~GHz. }
\label{f:vla}
\end{figure*}

Stars of masses $>$8~\msol\ spend a significant fraction of their
lifetimes, $>$10\%, embedded in their natal molecular clouds.
Single-dish (sub)millimeter observations have clarified the structure
of high-mass protostellar envelopes on $\sim$10$^4$--10$^5$~AU scales
(see \citealt{hvdt03} and references therein). However,
the distribution and kinematics of material on $\ltsim$1000~AU scales
is poorly known, due to the large ($\gtsim$1~kpc) distances involved,
and the lack of tracers at optical and near-infrared wavelengths.
These scales are of great interest to decide between formation
mechanisms for high-mass stars, and to clarify the relation with
clustered star formation \citep{evans02}. Does the distribution of
stellar masses in a star-forming region depend on the stellar density?
Also, the origin of the observed outflows and their interaction
with the environment on $\ltsim$1000~AU scales remain unclear.
Subarcsecond resolution observations are necessary to shed light on
these and other questions, which, at (sub)millimeter wavelengths, are
just coming within reach \citep{beuther04}.  However,
these resolutions can already be achieved in both the infrared and
radio wavebands, where extinction is much smaller than in the optical.

In the infrared, high-resolution techniques are most advanced at
near-infrared wavelengths.  Such observations probe less embedded,
more evolved phases, where a significant part of the surroundings is
already ionized.  Important progress has been made with the
identification of the ionizing stars of several ultracompact \hii\ 
regions (e.g., \citealt{watson+hanson}, \citealt{feldt03}).  In the
case of intermediate-mass stars, the imaging of the hot inner regions of disks
is presently generating a lot of interest
(\citealt{danchi01,tuthill01}).  In addition, a few more embeddded
objects have been probed (\citealt{weigelt02,preibisch02}), although
in the general case long baseline interferometers will be needed to
tackle most targets given the characteristic size scales involved
\citep{monnier02}.  At mid-infrared wavelengths, pioneering work has
been done by \citet{walsh01}, but subarcsecond resolution has only
recently been achieved (\citealt{tuthill02,buizer02,greenhill04}).

In the cm-wave region, most subarcsecond-resolution studies have
concentrated on \hho\ masers, which are bright
%Although masers do not trace the bulk of the material, and non-
%detections of masers are difficult to interpret, the emission is strong
enough for Very Long Baseline Interferometry (VLBI) observations. 
The excitation requirements of the masers are such that the emission
usually traces shocks associated with infalling or outflowing motions.
The VLBI data indicate that the maser emission traces moving gas
parcels, rather than shock waves propagating through an \hho--rich
cloud.  
In the case of outflow motions, both bipolar
and spherical flows are seen, which may represent different stages of evolution
\citep{torrelles03}.
In some cases, \hho\ masers in star-forming
regions may arise in accretion shocks in infalling gas \citep{mvdt04}.

Continuum emission at centimeter wavelengths arises in ionized gas.
In stellar winds and outflows, the gas can be collisionally ionized,
and VLBI data indicate a mixture of bipolar and equatorial outflows
\citep{hoare02}. Close to hot stars, small regions of photo-ionized
gas are observed as `hypercompact' \hii\ regions, which represent a
very early stage of high-mass star formation \citep{garay99}.

At a distance of 1.83$\pm$0.14~kpc \citep{imai00}, W3~IRS5 is the
nearest region of high-mass ($L$=1.2$\times$10$^5$~\lsol:
\citealt{ladd93}) star formation after Orion. The bright mid-infrared
source has been resolved into a double (\citealt{Howell81,Neugebauer82}). Single-dish
submillimeter mapping indicate an envelope mass of 262~\msol\ within a
radius of 60,000~AU, with an $r^{-1.5}$ density distribution
\citep{fvdt00}.  Near-infrared imaging shows a dense cluster
($\sim$3000 pc$^{-3}$: \citealt{megeath96}), mostly composed of
low-mass pre-main-sequence stars with ages 0.3--1~Myr \citep{ojha04}.
Radio continuum observations show a cluster of at least six
`hypercompact' \hii\ regions labeled A...F (\citealt{claus94,achim97}),
at least one of which exhibits proper motions \citep{wilson03}. Water maser
mapping reveals $\sim$100 spots, grouped in two flows: one roughly
spherical and centered close to continuum source~A, and the other more
collimated and centered close to source~D (\citealt{claus94,imai00}).
Mid-infrared spectroscopy shows CO absorption features blueshifted by
4..46~\kms\ relative to the systemic velocity \citep{mitch91}, which
must arise in an outflow. In millimeter-wave CO emission, blue- and
redshifted outflow lobes are detected out to $\approx$23~\kms\ from
the systemic velocity \citep{claussen84}, indicating that the
highest-velocity gas is very compact.
Finally, \emph{Chandra} observations by \citet{hofner02} indicate an
X-ray luminosity of W3~IRS5 of $L_X$=9$\times$10$^{29}$ erg~s$^{-1}$
(for $d$=1.83~kpc), consistent with the typical values for T~Tauri
stars.

This paper presents new cm-wave and mid-infrared images of W3~IRS5 at
sub-arcsecond resolution. The goals are to clarify the nature of the
radio continuum sources and their relation with the infrared double,
to find which ones are self-luminous, and which ones power the region.

\section{Observations}

\begin{figure*}[t]
\includegraphics[width=16cm]{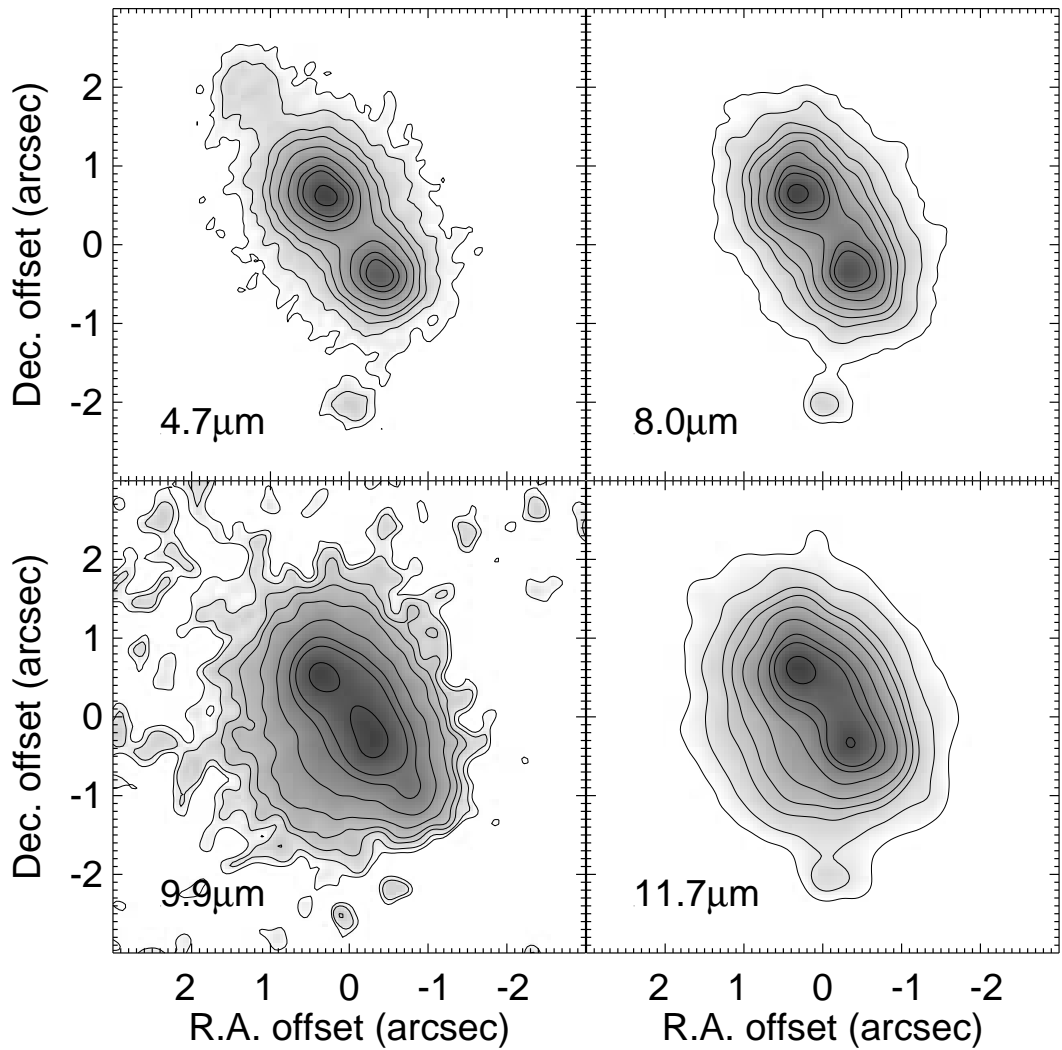}
\caption{Mid-infrared long-exposure images of W3 IRS5. Contours are at 
  0.5, 1, 2, 3, 5, 10, 20, 30 and 70\% of the peak intensity.}
\label{f:keck_long}
\end{figure*}

\begin{figure*}[t]
\includegraphics[width=16cm]{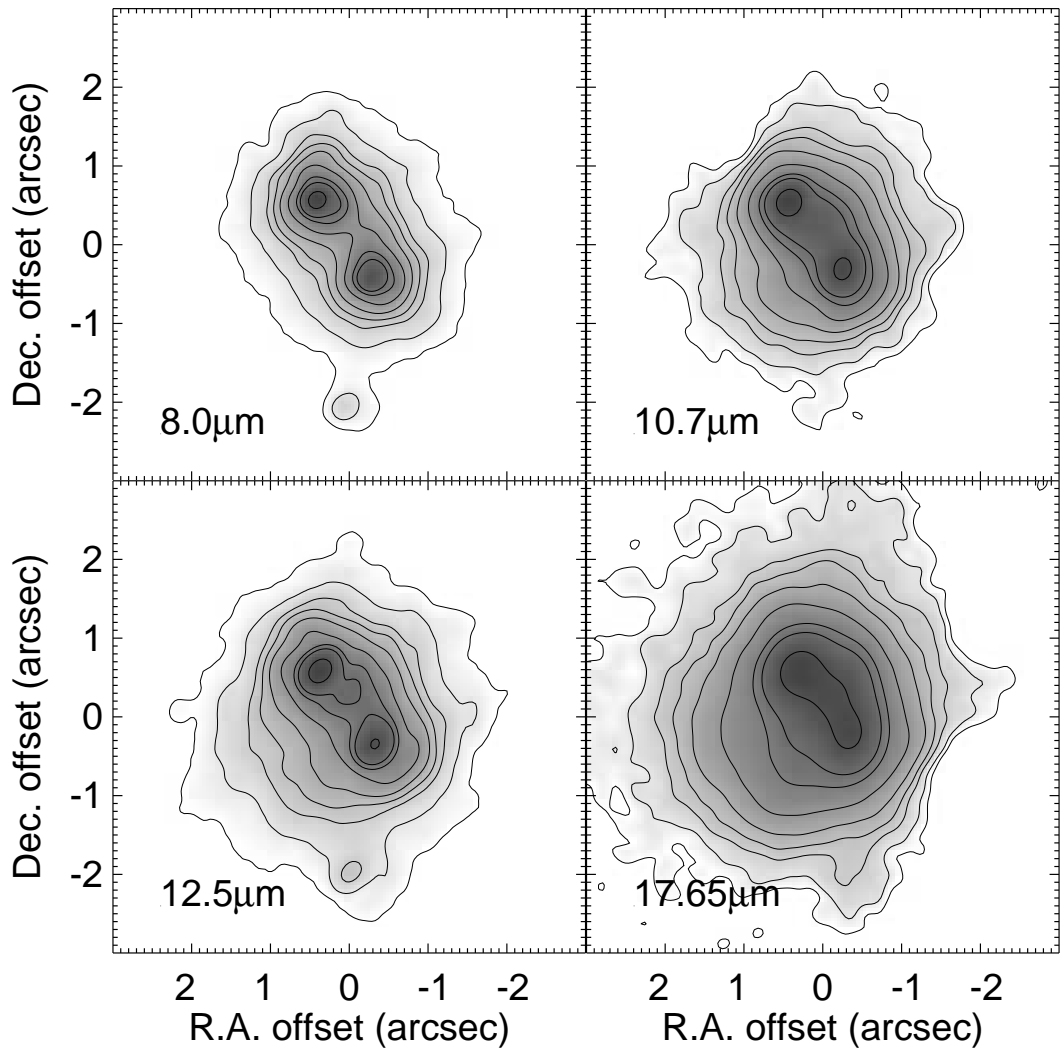}
\caption{Mid-infrared `speckle' images of W3 IRS5. Contours are at 
  0.5, 1, 2, 3, 5, 10, 20, 30 and 70\% of the peak intensity.}
\label{f:keck_spec}
\end{figure*}

\subsection{Radio Observations}

Radio observations of W3 IRS5 were carried out with the
NRAO\footnote{The National Radio Astronomy Observatory (NRAO) is
  operated by Associated Universities, Inc., under a cooperative
  agreement with the National Science Foundation.}  Very Large Array
(VLA) on 1996 October 24, when the VLA was in its $A$-configuration.
At this time, thirteen VLA antennas were equipped with 43~GHz
receivers; the other fourteen observed at 22~GHz (respectively 
known as \textit{Q}- and \textit{K}-band in radio astronomy).
 Zenith opacity was
0.089 at 22.5~GHz and 0.071 at 43.3~GHz.  Elevation-dependent antenna
gains were interpolated from values measured by the VLA staff.  The
phase calibrator, 0228+673, was observed every 10 minutes at 43~GHz
and every 15 minutes at 22~GHz (the `fast switching' procedure was not
implemented at the time). Pointing was checked at 8.4~GHz on the same
source every 70 minutes at 43~GHz and every 5~hours at 22~GHz.
On-source integration time is 438~min at 22~GHz and 400~min at 43~GHz.
The data were edited, calibrated, and imaged with NRAO's Astronomical
Image Processing System (AIPS). Absolute calibration was obtained from
observations of 3C~286 using flux densities interpolated from the
values given by \citet{ott94}. 
For 0228+673, we obtain a flux density of 1.83~Jy at 22~GHz and
1.55~Jy at 43~GHz.

Figure~\ref{f:vla} shows the 22~and 43~GHz maps, which have rms noise
levels of 0.15 and 0.16 mJy~beam$^{-1}$. These maps were obtained from
the $uv$ data by a Fourier transform with uniform weighting, and
deconvolved with the Clean algorithm.  Restoring beam major and minor
axes and position angles are 44$\times$37 mas at position angle
--68$^\circ$ at 43~GHz and 89$\times$88 mas, PA --53$^\circ$ at
22~GHz.

\subsection{Infrared Observations}

Data were obtained in August 2002 with the Long Wavelength Spectrometer
(LWS) camera on the Keck~I telescope\footnote{The W.~M.\ Keck Observatory 
  was made possible by the support of the W.~M.\ Keck Foundation, and is 
  operated as a scientific partnership among the California Institute of 
  Technology, the University of California, and the National Aeronautics 
  and Space Administration.}. 
Two different observing methodologies were employed. 
The first set of observations utilized a standard chop-nod pattern, 
with frames coadded to build up longer exposure times.
Observations of W3 IRS5 (with the calibrator $\xi$ Cyg) taken in this 
mode emphasized the recovery of faint structure.
The second mode had a fast readout in addition to the chop-nod, so that
large volumes of rapid exposure data were collected in a fashion 
analogous to a mid-IR `speckle' experiment, with the hope to 
recover fine structure in the images. Again, these objects
were paired with similar data taken on a point source calibrator, this
time $\alpha$ Ceti. No point-source calibrator files were taken for
9.9~\mic\ or 12.5~\mic\ observations. 
These two modes are denoted `L' and `S' for `long' and `speckle' exposure
hereafter. 

Data have been analyzed with an iterative matched filter version of
the shift-and-add algorithm. This algorithm attempts to match the
shifts in the current iteration to maximise the correlation with the
output of the previous iteration. Significant gains in image resolution
were demonstrated over straight coadding, while a simple shift-and-add
strategy was foiled in this case by the presence of two nearly equal
peaks.

Figures~\ref{f:keck_long} and~\ref{f:keck_spec} present the resultant
images, except the 17.65~\mic\ long-exposure image which is very
similar to the speckle image at that wavelength. Note that an artifact
due to an unwanted reflection from an optical surface affects the
4.7~\mic\ image, giving a spurious feature to the North-East which is
not seen at any other wavelength. This `ghost' was also present in
point-source calibrator data in this filter.  The 9.9~\mic\ and (to a
lesser extent) the 10.7~\mic\ data have degraded signal-to-noise due
to the much lower flux levels.  There were, unfortunately, additional
experimental difficulties which were not easy to account for.  The
observations were taken under conditions of variable cirrus,
increasing the errors in photometry.  Compounding this was an
intermittent mechanical fault with the camera mechanism which resulted
in partial occultation of the pupil, affecting both the throughput and
the point-spread function (PSF).  Although this had little effect on
the maps presented here, it did preclude our original intent of fully
deconvolving the images using the PSF from the reference star
observations.

\section{Results}

\begin{table*}
\caption{Radio emission from W3 IRS5. Numbers in brackets are
  uncertainties in units of the last decimal.}
\label{t:vla}

\begin{tabular}{llllllll}

\hline
\hline

Source & $\alpha$ & $\delta$ & Peak $I_\nu$ & Total $S_\nu$ & Major axis & Minor axis & PA \\
       & (J2000)  & (J2000)  & mJy/beam     & mJy           & mas        & mas        & deg \\
\hline
 Q1 & 02 25 40.660407(635)  & 62 05 51.82215(404)  & 0.691(162) & 0.78(31) & 45(10)& 41(10)& 93(90) \\
 Q2 & 02 25 40.676344(379)  & 62 05 52.04937(261)  & 1.285(158) & 1.92(36) & 59(7) & 41(5) & 47(13) \\
 Q3 & 02 25 40.681511(211)  & 62 05 51.53000(198)  & 2.102(155) & 3.89(42) & 63(5) & 47(3) & 7(10)  \\
 Q4 & 02 25 40.728475(704)  & 62 05 49.85180(702)  & 0.666(154) & 1.43(46) & 75(17)& 46(11)& 21(18) \\
 Q5 & 02 25 40.783441(212)  & 62 05 52.46552(141)  & 1.996(163) & 2.19(30) & 43(4) & 41(3) & 88(57) \\
    & & & & \\                                                                          
 K1 & 02 25 40.143241(1234) & 62 05 52.28144(1080) & 0.745(142) & 0.85(27) & 139(27) & 99(19) & 25(22) \\
 K2 & 02 25 40.661233(1029) & 62 05 51.86622(1174) & 0.847(140) & 1.08(29) & 175(29) & 88(15) & 160(9) \\
 K3 & 02 25 40.675758(1135) & 62 05 52.04087(690)  & 0.857(142) & 0.76(23) & 121(20) & 88(15) & 59(21) \\
 K4 & 02 25 40.682172(526)  & 62 05 51.54376(441)  & 1.938(139) & 2.76(31) & 152(11) & 113(8) & 154(10) \\
 K5 & 02 25 40.712459(785)  & 62 05 52.44786(569)  & 1.163(142) & 1.12(24) & 110(13) & 106(13)& 168(138)\\
 K6 & 02 25 40.733047(1601) & 62 05 49.84977(1214) & 0.797(134) & 1.76(41) & 174(29) & 154(26)& 25(56) \\
 K7 & 02 25 40.782747(737)  & 62 05 52.46103(637)  & 1.313(140) & 1.63(28) & 150(16) & 100(11)& 29(10) \\
 K8 & 02 25 40.861211(888)  & 62 05 53.46070(717)  & 1.307(135) & 2.44(37) & 169(17) & 135(14)& 25(19) \\

\hline
\end{tabular}
\end{table*}

\subsection{Radio Positions}
\label{s:cm_pos}

Table~\ref{t:vla} lists the sources detected above 5$\sigma$ using the
multiple-peak-finding AIPS task SAD after trying various weighting
schemes. 
For the $Q$--band data, the table uses the image obtained with uniform
weighting, which has better positional accuracy, while for the $K$--band data,
tabulated results use an image obtained with natural weighting, where more
sources are detected (rms = 142~$\mu$Jy).  
Sources Q1 and Q4 are 4$\sigma$ detections and were not found by SAD,
but their detection is secure because they have $K$--band
counterparts.
The flux densities in the table are corrected for primary beam
response. Positional uncertainties are statistical errors and apply to
the relative positions of these radio sources. 

Columns 6--8 of Table~\ref{t:vla} gives the source sizes.  Since only
sources Q3, K6 and K8 appear marginally resolved, the sizes have not
been deconvolved. The sizes would be lower limits if extended emission
is resolved out by the interferometer (\citealt{kurtz99};
\citealt{kim01}). In the particular case of W~3 IRS5, however,
multi-configuration observations rule out extended emission down to
very low limits \citep{achim97}. Therefore, the radio sources of
W3~IRS5 belong to the class of `hypercompact' \hii\ regions
\citep{kurtz02}.

Comparing the positions in Table~\ref{t:vla} with those from 1989 (\citealt{claus94};
\citealt{achim97}) leads to the following identifications: Q1 = K2, Q2 = K3 = A, Q3 = K4 =
B, Q4 = K6, Q5 = K7 = MD1, K5 = C, K8 = F. Source K1 is several arc seconds
away and probably unrelated. Sources Q1=K2 and Q4=K6 were not seen before and appear to be
new. On the other hand, sources E and G seem to have disappeared since 1989,
the epoch when the Tieftrunk et al data were taken. Most strikingly,
source D2 has disappeared, which is remarkable since it was
the strongest source in 1989. Perhaps D1 and D2 were not separate sources, but
merely substructure within one source, which we refer to as D hereafter.

Our sources with counterparts in the old data do not exactly lie on the
positions reported by \citet{claus94} and \citet{achim97}, but rather at
80--140 mas shifts. The shifts are 2--3 beam sizes and $>$10 times the formal
error on relative positions.  The position angles of the shifts vary between
20$^\circ$ and 60$^\circ$, which argues against instrumental effects such as
pointing errors or changes in calibrator positions, which would shift all sources in
the same direction.  The data thus seem to confirm the existence of proper motions
reported by \citet{wilson03}.  At a distance of 1.83~kpc, a motion of 100~mas
in 7.48~yr corresponds to a transverse velocity of 116~\kms.  The space
velocity may be a factor $\sqrt{2}$ higher, or 164~\kms.  These values are
much larger than the motions of the \hho\ masers of $\sim$20 \kms\ 
\citep{imai00} and of the CO emission and absorption (\S~\ref{s:intro}).

\subsection{Infrared Positions}
\label{s:ir_pos}

\begin{table*}
\caption{Quantitative analysis of the infrared imaging: flux densities 
and relative positions. The table columns are arranged as follows:
(1) observing $\lambda / \Delta \lambda$; (2) speckle `S' or long `L'
exposure; (3) FWHM of point-source reference star; (4,6,10) flux of 
MIR 1--3; (5,7,11) FWHM of MIR 1--3; (8,12) separation of 
sources MIR1-2 and MIR1-3; (9,13) position angle of sources MIR1-2 and MIR1-3;
(14) total flux detected. }
\label{t:ir}
\begin{tabular}{rccrrrrrrrrrrr}
\hline
\hline
Filter & Mode & PSF star & 
\multicolumn{2}{c}{MIR1}&
             \multicolumn{2}{c}{MIR2}& \multicolumn{2}{c}{MIR1-MIR2}&   
               \multicolumn{2}{c}{MIR3}& \multicolumn{2}{c}{MIR1-MIR3}& Total \\
$\lambda / \Delta \lambda$ 
          &L/S&FWHM & flux& FWHM& flux& FWHM& Sep  & PA  & flux & FWHM& Sep  & PA  & Flux  \\
($\mu$m)  &   &(mas)& (Jy)&(mas)& (Jy)&(mas)& (mas)&(deg)& (Jy) &(mas)& (mas)&(deg)& (Jy)  \\
\hline
  4.8/0.6 & L & 507 &  44 & 453 &  23 & 471 & 1217 & 215 &  0.6 & 480 & 2730 & 186 &  80 \\
  8.0/0.7 & L & 384 & 140 & 432 & 108 & 480 & 1198 & 215 &  2.0 & 436 & 2741 & 187 & 301 \\
  8.0/0.7 & S & 258 & 152 & 310 & 123 & 367 & 1214 & 217 &  1.8 & 304 & 2663 & 188 & 328 \\
  9.9/0.8 & L &  -  &   4 & 576 &   4 & 568 & 1013 & 215 &$<0.5$& ~-~ & ~-~~ & ~-~ &  11 \\
 10.7/1.4 & S & 291 &  13 & 460 &  12 & 552 & 1078 & 219 &$<1.2$& ~-~ & ~-~~ & ~-~ &  32 \\
 11.7/1.0 & L & 364 &  60 & 519 &  50 & 607 & 1097 & 216 &  0.7 & 255 & 2739 & 187 & 142 \\
 12.5/0.9 & S &  -  &  60 & 467 &  50 & 587 & 1145 & 218 &  1.1 & 329 & 2632 & 188 & 146 \\
 17.65/0.9& L & 451 &  98 & 716 &  92 & 636 & 1049 & 217&$<10.7$& ~-~ & ~-~~ & ~-~ & 318 \\
 17.65/0.9& S & 452 & 107 & 646 &  93 & 615 & 1102 & 219&$<10.3$& ~-~ & ~-~~ & ~-~ & 323 \\
\hline
\end{tabular}
\end{table*}

The infrared images (Figs.~\ref{f:keck_long} and~\ref{f:keck_spec})
show three compact sources, surrounded by diffuse emission which becomes
more pronounced toward longer wavelengths.  We begin a quantitative
analysis of these images by fitting simple profiles to the data, and
by measuring flux densities in different regions.  The results are
given in Table~\ref{t:ir}, which gives the fluxes, relative positions
and full-width at half-maximum (FWHM) of the various components.  In
addition, the FWHM of the point-source reference stars are given,
which gives an estimate of the resultant system PSF.  Examination of
these data shows that the system appears truly diffraction-limited in
either mode (`L' or `S') at 17.65~\mic.  At shorter wavelengths
(8.0--12.5~\mic), the `S' mode delivers a significantly smaller FWHM
than `L' (8.0~\mic\ gives a direct comparison) which approaches the
formal diffraction limit.  The dramatic increase in size at 4.8~\mic\ 
implies some optical problem beyond the normal effects of diffraction
and seeing, such as optical aberration or defocus.

In the absence of a wide-field image with standard stars, our only astrometric
information comes from the relative positions of the components.  We refer to
the northernmost bright component as MIR1, with MIR2 being of nearly equal
brightness to the south, and MIR3 the much fainter southernmost peak.
Table~\ref{t:ir} lists the separation and the position angle of MIR2 \& MIR3
relative to MIR1 for all observations, obtained through Gaussian fits to the
emission.

The mean separation of MIR1 \& MIR2 is 1124$\pm$74 mas at a position angle of
36.8$\pm$1.7 degrees.  These values are consistent with those from earlier
mid-infrared work (\citealt{Howell81,Neugebauer82}), but our data are the
first to image the mid-infrared double directly.

We compare this relative position with those of pairs of radio
sources. The best match is for pair Q3--Q5, whose separation of 1210~mas at
a position angle of 37\fdeg4 is in good agreement with the infrared peaks.
The only other radio pair with similar relative positions is K7--K8, which has
a separation of 1141 mas, but at a position angle of 28\fdeg9, inconsistent
with the infrared result.  On this basis, we identify the bright mid-infrared sources
with radio sources Q5=K7=MIR1 and Q3=K4=MIR2.  
%Radio sources K2, K3 and K5, which lie in between Q3 and Q5, may be
%associated with the diffuse infrared emission.
The position of source MIR3, with 1--2\% of the flux density of the
main sources, then coincides with radio source Q4=K6, confirming our
identification. There is a cluster of \hho\ maser spots close to this
object, at $\Delta$($\alpha$)$\approx$250 and
$\Delta$($\delta$)$\approx$2000 mas \citep{imai00}. 
Table~\ref{t:id} summarizes our source identifications.

The images in Figures~\ref{f:keck_long} and~\ref{f:keck_spec} show the only
region of flux detected within the 10\farcs24 field of view of LWS, with one
exception. In a few frames (which happened to be offset from center) a faint
diffuse component was seen at the extreme edge of the field, 7\farcs3 from MIR1 
at a position angle of 160$^\circ$.  Using the radio identifications of MIR1 and
MIR2 as astrometric reference, this source, which we call MIR4, lies at position
$\alpha$ = 02$^h$ 25$^m$ 41\fsec1388, $\delta$ = 62$^\circ$ 05$'$ 45\farcs604
(J2000), where no radio emission is detected. The extended nature and
location at the edge of the field of view preclude measurement of its
mid-infrared flux density.

\begin{table}
\caption{Radio and infrared identifications.}
\label{t:id}
\begin{center}
\begin{tabular}{cccc}

\hline
\hline

MIR & Q & K & 1989 \\

\hline

1 & 5 & 7 & D \\
2 & 3 & 4 & B \\
3 & 4 & 6 & --\\
--& 1 & 2 & --\\
--& 2 & 3 & A \\
--& --& 5 & C \\
--& --& 8 & F \\
--& --& --& E \\
--& --& --& G \\
--& --& 1 & -- \\

\hline
\end{tabular}
\end{center}
\end{table}

\subsection{Radio Brightness}
\label{s:cm_b}

\begin{figure}[b]
  \begin{center}
\includegraphics[width=6cm,angle=-90]{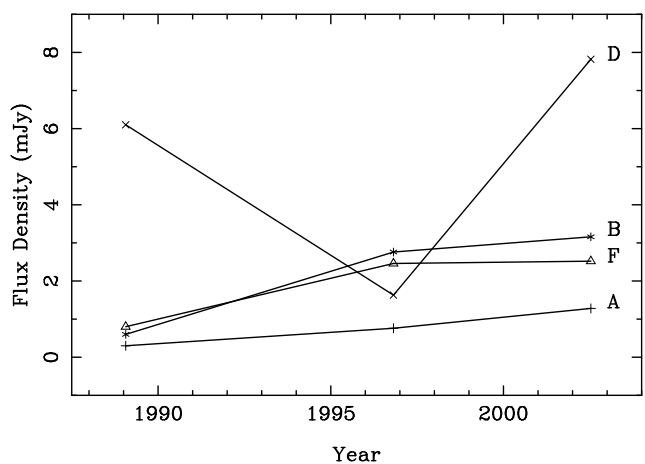}
\caption{Flux densities (at 22 GHz) of radio sources in W3 IRS5 as a function of time.
         Data are from \citet{achim97}, \citet{wilson03}, and Table~\ref{t:vla}.}
\label{fig:fdevol}
\end{center}
\end{figure}

The flux densities of 22~GHz sources K3, K4, K5, K7 and K8 are
significantly different from the values by \citet{achim97}, probably
due to variability. Instrumental effects, such as calibration
problems, atmospheric decorrelation, or difference in beam size, would
affect all sources in the same way. Instead, several components seem
to undergo gradual increases or decreases in 22~GHz brightness over
the available 13-year period (Figure~\ref{fig:fdevol}).

The total flux densities in Table~\ref{t:vla} indicate a spectral
index $\gamma$, defined through 
$$S_\nu \propto \nu^\gamma$$ 
of $\gamma$=1.42$\pm$0.77
for source Q2, $\gamma$=0.52$\pm$0.33 for source Q3, and
$\gamma$=0.45$\pm$0.48 for source Q5. Values derived from the peak
brightness are 0.62$\pm$0.44, 0.12$\pm$0.22 and 0.64$\pm$0.29. These
values are consistent with thermal emission, and are not affected by
variability as the data were taken simultaneously. For the weak 43~GHz
sources Q1 and Q4, we find $\gamma$=--0.3$\pm$0.6, i.e., flat or
slightly nonthermal spectra. Sources K1, K5 and K8, which are not
detected at 43~GHz, have $\gamma$=--1$\pm$1 and may be of nonthermal
nature. The spectral indices found here are consistent with those by
\citet{wilson03}, which are also based on simultaneous measurements,
except for source K8=F which was detected at 43~GHz in 2002 but not in
1996.

In the case of Bremsstrahlung from an ionized region with a power law
distribution of the electron density with radius, 
$$n_e \propto r^{-q}$$
the spectral index is 
$$\gamma = (2q-3.1)/(q-0.5)$$
\citep{olnon75}.  Hence the spectral index of $\gamma \approx 0.5$
measured for Q3 and Q5 corresponds to $q\approx1.9$, close to the
value of $2$ for an ionized wind.  Observations of broad H~I radio
recombination lines (\citealt{achim97,sewilo04}) support this
interpretation, although higher angular resolution is needed to relate
the line emission with the continuum sources.  In any case, our
measurements are not sensitive enough to rule out a flat radio
spectrum for these sources, which would indicate optically thin
emission. The objects could then be externally ionized, consistent
with the absence of mid-infrared emission.

The value $\gamma\approx 1.5$ measured for Q2 could arise in an \hii\ 
region with a constant density at the center and a steep
($q\approx4.7$) outer fall-off.  It is not clear which mechanism would
create such a density distribution, especially since the sound speed
of $v_S \sim$10~\kms\ of \hii~regions implies that in these compact
(\O $\ltsim$100~AU) sources, density fluctuations are washed out
within $\sim$50~yr. Therefore, Q2 is probably a uniform-density
\hii~region which is (moderately) optically thick, but again, the data
do not rule out a wind spectrum.  In either case, it is internally
ionized.

More sensitive measurements over a larger wavelength range are
necessary to constrain the emission mechanism. A constant spectral
index would support the wind model, while a bent spectrum would
indicate uniform \hii~regions of intermediate optical depth.
%% Multiwavelength observations were performed by \citet{achim97}, who
%% found a mixture of thermal and nonthermal emission. This result is
%% hard to understand, and may be due to source variability.
Care has to be taken, however, not to include dust emission at high
frequencies, or synchrotron emission at low frequencies
(\citealt{felli93}; \citealt{reid95}).

\subsection{Infrared Brightness}
\label{s:ir_bright}

Columns 4, 6 and 10 of Table~\ref{t:ir} report the flux densities of the three
mid-infrared sources.  Brightness was measured in circles of radius 600 mas,
which cover all of the diffraction and seeing patterns.  For the purposes of
flux calibration, six reference stars were used, and four of these gave
consistent photometric results (the other two, presumably affected by
non-photometric conditions, were ignored).  At 9.9 and 12.5~\mic\ where 
there were no calibrator star data taken in an identical way, the
flux calibration was derived from indirect measurements and should be
regarded as tentative. Due to these difficulties and the
variable vignetting in the camera, we quote errors on the photometry of up to
50\%.
Within these errors,
the total flux densities at 4.8--10.7~\mic\ are consistent with the values
measured by \citet{willner82} in a $\sim$10$''$ beam.
However, at 9.9~and 12.5~\mic, where no calibrators were observed, and
at 11.7~\mic, where only one calibrator was observed, the photometric
error is probably closer to a factor of 2. Indeed, at 11.7 and
12.5~\mic, our flux densities are a factor of $\sim$2 below the values
measured by Willner et al and by \citet{persi96} in a 3$''$ beam.

It is interesting to note that the relative fluxes and positional offsets
between MIR1--3 remain fairly constant across the mid-infrared and (presuming
our radio identifications) into the radio.  This implies that it is unlikely
that there are large differences in the effective temperatures or the optical
depths to these 3 components.  The only readily identifiable exception to this
is a systematic trend for MIR1 being brighter than MIR2 at short wavelengths,
while they are nearly equal at long wavelengths.  This may imply a somewhat
hotter underlying spectrum, or there may be opacity gradients in the line of
sight.

\subsection{Contribution from PAHs}
\label{s:pah}

The observed infrared emission may be continuum emission from dust grains.
However, the mid-infrared spectra of many Galactic sources, including compact
\hii\ regions and other star-forming regions, show strong emission
features due to Polycyclic Aromatic Hydrocarbons (PAHs).
The strongest PAH features lie at 3.3, 6.2, 7.7, 8.6, 11.2 and
12.7~\mic\ \citep{peeters02}. Therefore our 8~\mic\ filter contains the 7.7~\mic\
feature, the 10.7 and 11.7~\mic\ filters the 11.2~\mic\ feature, and
the 12.5~\mic\ filter the 12.7~\mic\ feature. 
The strength of these features reflect the ambient UV radiation field,
rather than the dust mass or temperature.
Therefore, to interpret our Keck data properly, quantifying the
contribution of PAHs to the observed emission is essential.

We have searched the ISO-SWS spectrum of W3~IRS5 (F.~Lahuis, priv.\
comm.) for PAH features. With typical widths of 0.1--0.4~\mic, the
features should be easily resolved with ISO. No PAH features are
detected down to an rms noise level of $\approx$1~Jy. The flux
densities measured with ISO are 60--90\% of those measured with Keck,
so beam dilution does not play a role. We conclude that the emission
observed with Keck is thermal emission from dust grains. 

\subsection{Infrared Sizes}
\label{s:ir_size}

Our fits to the mid-infrared images show that the two bright cores
MIR1 and MIR2 exhibit systematically larger sizes than the reference
stars, as measured by the Gaussian FWHM values given in Column~3 of
Table~\ref{t:ir}.  In this section, we extract quantitative estimates
of the apparent angular diameters of these cores, by deconvolving with
the reference star PSF then fitting with a simple circularly-symmetric
profile (in this case a uniform disk).

However, this could only be done in a minority of cases where the
data were suitable and of sufficient quality.
The 4.8~\mic\ data were affected by an unknown optical problem (as
discussed earlier), while the 9.9 \& 12.5~\mic\ data had no PSF 
reference star data, and were ignored here.
Furthermore, deconvolution requires the highest possible angular 
resolution data, and we therefore restrict our attention to only the 
rapid exposure observing mode `S', discarding `L' (e.g. all 11.7~\mic\ 
data).
Perhaps the most difficult aspect of the deconvolution problem was
distinguishing between the resolved cores and the more extended nebula.
At the longest wavelength, 17.65~\mic, this was not possible for
two reasons: firstly the extended component was relatively bright
compared to the cores, and secondly the angular resolution was not
sufficient to clearly distinguish between compact and extended flux.

The remaining datasets suitable for deconvolution and diameter fitting
were from 8.0 \& 10.7~\mic. Fits were obtained with a uniform circular
disk profile, although in the partially resolved case as here any
simple model (such as a Gaussian) would serve equally well.  Uniform
disk fits were obtained for MIR1 \& MIR2 at 8.0~\mic, where the
diameters are 207 \& 254~mas, and at 10.7~\mic, where they are 300 \&
333~mas, respectively.  Although the formal errors on these quantities
are around 40~mas, the true uncertainties are hard to quantify due to
unknown seeing and optical changes between the source and calibrator
stars, and due to imperfect rejection of the extended nebula in
fitting the core.

Two systematic trends are noted here: MIR2 appears slightly larger than MIR1,
and the sizes at 10.7~\mic\ are larger than those at 8.0~\mic.  However,
particular caution needs to be expressed over contamination from the extended
flux component which may have a role in causing these apparent extensions.

The measured mid-infrared sizes exceed the limits on the radio sizes
of $\sim$100~AU (Table~\ref{t:vla}). This result supports a model
where the radio emission comes from ionized gas very close to a star
and the mid-infrared emission from warm dust somewhat further out.

\section{Discussion}
\label{disc}

Our observational findings can be summarized as follows: three
mid-infrared point sources with radio counterparts; four radio sources
without mid-infrared counterparts that appear to change position; and
diffuse mid-infrared emission.
The following sections discuss each of these components in turn.

\subsection{Mid-infrared point sources}
\label{sec:blackbody}

\begin{figure}[tb]
\includegraphics[width=6cm,angle=-90]{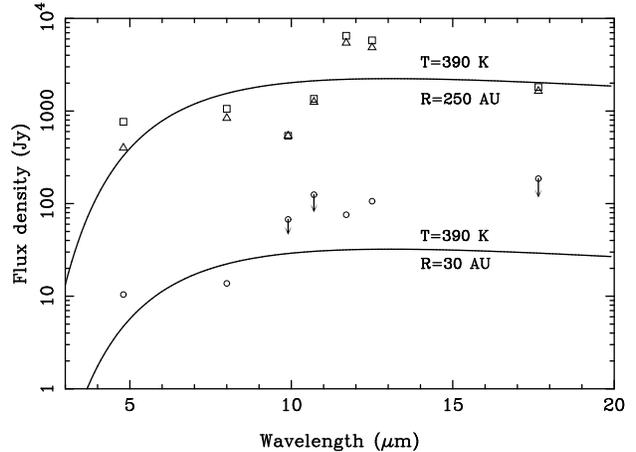}
\caption{Mid-infrared spectral energy distributions of W3 IRS5 after
  dereddening. Squares, triangles and dots indicate
  MIR1, MIR2 and MIR3. Model curves are superposed.}
\label{f:dered}
\end{figure}

To estimate the physical properties of the compact mid-infrared sources, we have
compared their flux densities to a simple blackbody model. Given the lack of
strong observed colour variations (\S~\ref{s:ir_bright}), we assume that the
sources have the same temperature and foreground extinction.  Based on
the results of \S~\ref{s:ir_size}, we use a radius of 250~AU for sources MIR1
and MIR2; for MIR3, $R$=30~AU is adopted based on its lower brightness.  The
observed far-infrared luminosity then limits the temperature to $T<$390~K,
which appears plausible based on the measured colours and 
the 2.2~\mic\ photometry by \citet{ojha04}.

Using this temperature and radius, we can model the observed flux
densities if we know the foreground extinction.
The broad-band mid-infrared spectrum presented by \citet{willner82}
indicates a silicate optical depth of $\tau_S \approx$4.3--5.0
assuming pure absorption, or $\tau_S$=7.64 when correcting for
underlying silicate emission, as Willner et al do.
More recent data from ISO give consistent results (F.~Lahuis, priv.\ 
comm.), even though they refer to a larger beam ($\sim$20$''$) that
varies by a factor $\sim$4 over this wavelength range.

Figure~\ref{f:dered} shows the measured flux density spectrum after
de-reddening by $\tau_S$=5.0. Values for the extinction at other wavelengths
are computed using dust properties by \citet{oh94}, Model~5. This dust model
gives a good match between envelope masses derived from dust and CO
\citep{fvdt99}.  The figure shows that a blackbody model with $T$=390~K
reproduces the data within a factor of 3. The largest outliers are the
9.9~and 11.7~\mic\ points, at which wavelengths the calibration is the
most uncertain. 

The \citet{oh94} dust model has Si:C=1.45, while values up to $\approx$2 are
observed \citep{kruegel}. Increasing the Si:C ratio would improve the
match between data and model at 9.9~\mic, but would give a worse fit
at 11.7~\mic. More likely, the assumption of blackbody emission is not
quite valid, so that towards shorter wavelengths, smaller radii and
higher temperatures are probed. Geometrical effects may also influence
the shape of the silicate absorption.

Temperatures below 390~K are energetically allowed, but require lower
extinctions to fit the observed flux densities. For $T<350$~K, the
required extinction drops below $\tau_S$=4.3, which we consider
unlikely based on the \citet{willner82} data.  The total luminosity
for $T$=350~K is $\sim$8$\times$10$^4$~\lsol, which leaves
$\sim$4$\times$10$^4$~\lsol\ for a third power source, such as radio
source Q2 (\S~\ref{s:cm_b}). 
This option is more likely than the case of two
power sources, because radio sources Q2, Q3 and Q5 are of similar
strength (\S~\ref{s:cm_b}).

The (tentative) size increase of MIR1~\&~2 from 8.0 to 10.7~\mic\ 
(\S~\ref{s:ir_size}) is to be expected if a more realistic assumption of a
centrally-heated dust cloud with a thermal profile is adopted, rather than a
blackbody at a single temperature.  A detailed understanding of this deeply
embedded and complex region will clearly require radiative transfer modelling,
and it is encouraging that mid-infrared imaging appears capable of placing
meaningful constraints.

The best-fit extinction of $\tau_S$=5.0 is significantly below Willners
estimate.  Perhaps their simple formula to correct for silicate emission is
not valid at large extinction values. More likely, the silicate absorber is
physically decoupled from the underlying continuum emitter.  One geometrical
interpretation is that of two star/disk systems surrounded by a cavity, whose
walls cause the silicate absorption. Such a cavity would also explain why
submillimeter imaging in a 15$''$ beam indicates a much higher extinction
($A_V\sim$300; \citealt{fvdt00}) than the mid-infrared data.

\subsection{Stationary radio sources}
\label{sec:ionize}

The flux densities $S_\nu$ of the radio sources (Table~\ref{t:vla})
can be used to estimate the Lyman continuum emission $N_L$ of their
ionizing sources, assuming that the \hii~regions are uniform and
isothermal (see, e.g., \citealt{tools}). This discussion concentrates
on sources Q2, Q3 and Q5 which have positive spectral indices
(\S~\ref{s:cm_b}), and considers both optically thin and optically
thick emission as limiting cases. In the optically thin case, $N_L$ is
directly proportional to the flux density. In the optically thick
case, black body emission at $T$=10$^4$~K indicates radii of
$\sim$20~AU, consistent with the observational upper limits
(Table~\ref{t:vla}). The emission measure follows from setting the
free-free optical depth equal to unity; radii and emission measures
together indicate electron densities of 10$^6$--10$^7$~\ccm. Balancing
photoionization with `case B' recombination \citep{osterbrock} finally
gives $N_L$. The results for both cases are
$N_L$=1...7$\times$10$^{44}$~s$^{-1}$, with a weak dependence on
electron temperature.

The similar values of $N_L$ derived for 43~GHz sources Q2, Q3 and Q5 indicate that
they have similar luminosities.  Therefore it is hard to see how only
the ionizing source of Q2 could be invisible in the mid-infrared,
unless it is extremely deeply embedded.  In fact, our 17.65~\mic\ 
images may show a source about 0\farcs5 North of MIR2, but the data do
not allow to extract a flux density.

The stellar luminosities of $\approx$40,000~\lsol\
(\S~\ref{sec:blackbody}) correspond to masses of $\approx$20~\msol\
and ZAMS spectral types O8 \citep{maeder89}.  Their expected Lyman
continuum emissions are $\approx$6$\times$10$^{48}$~s$^{-1}$
\citep{sdk97}, which is $\sim$10$^4$ times the value just derived from
the radio continuum emission. Since dust absorption inside the
\hii~region only accounts for factors of 2--3, this discrepancy may be
due to accretion of dust particles \citep{walmsley95}. The observed
variability (Fig.~\ref{fig:fdevol}) may then correspond to variations
in the accretion rate.

Accretion of dust would also explain why the hypercompact \hii\
regions stay confined to a $\sim$20~AU radius. The ionization front
around an O-type star on the main sequence is a D-critical front (e.g.,
\citealt{osterbrock}) which moves at about the sound speed of
$\approx$10~\kms, or somewhat less (5--7~\kms) since the surrounding
H~I shell needs to be accelerated.  \citet{acord98} have seen such
expanding motions in the ultracompact \hii\ region G5.89. In the case
of W3~IRS5, expansion at 5--10~\kms\ would lead to an increase in
radius from 20 to 100~AU over the observed 10-year period which is not
observed.

The accretion rate needed to confine the hypercompact \hii\ regions may
be estimated by equating the accretion force (momentum transfer rate)
of the dust to the thermal pressure of the \hii\ region.  Using a
radius of 20~AU, a density of 3$\times$10$^6$~\ccm, and $T$=8000~K for
the \hii\ region, we find $\dot{M}$=1.5$\times$10$^{-8}$~\msol\,yr$^{-1}$
assuming that the dust is in free fall onto a 20~\msol\ star. In
reality, radiation pressure will slow down the dust from the free-fall
speed (42.2~\kms), so that perhaps twice this $\dot{M}$ is needed.
If the stars have winds with substantial mass loss rates (e.g.,
10$^{-6}$~\msol\,yr$^{-1}$), even higher accretion rates may be
needed to confine the \hii\ region.

Recent work by \citet{keto02}, however, shows that stellar gravity
prevents the hydrodynamical expansion of \hii\ regions inside a
`gravitational radius'
$$ r_g = GM / 2c_s^2$$
where $G$ is the gravitational constant, $M$ the stellar mass, and
$c_s$ the sound speed of \hii\ regions of $\approx$10~\kms. 
For the bright mid-infrared sources in W3~IRS5, $M\approx$20~\msol\
(\S~\ref{sec:blackbody}) so that $r_g\approx$90~AU, consistent
with the observational limits (Table~\ref{t:vla}).

In Keto's model, both the ionized region close to the star and the
surrounding molecular gas have free-fall density profiles, $n\propto
r^{-1.5}$. At $r=r_g$, the accretion flow changes from molecular to
ionized. Such a density profile was indeed found for the molecular
envelope of W3~IRS5 by \citet{fvdt00} from submillimeter continuum and
line maps.

The expected flux density of a gravitationally bound \hii\ region only
depends on the density $n_0$ at the radius $r_m$ where the molecular gas
reaches its sound velocity \citep{keto03}. 
Taking $T$=30~K for the molecular gas, $r_m\approx 0.35$~pc.
For $d$=1.83~kpc, $M$=20~\msol\ and $T_e$=10$^4$~K, the observed flux
density of $\approx$1~mJy at 22--43~GHz is reproduced for
$n_0\approx$1$\times$10$^5$~\ccm. 
This estimate agrees to a factor of 5 with the value of
$n_0$=2$\times$10$^4$~\ccm\ derived by \citet{fvdt00}.
We conclude that gravitation explains the compactness of the radio
 sources in W3~IRS5 which have mid-infrared counterparts.

\subsection{Transient radio sources: proper motions?}

\begin{figure}[bt]
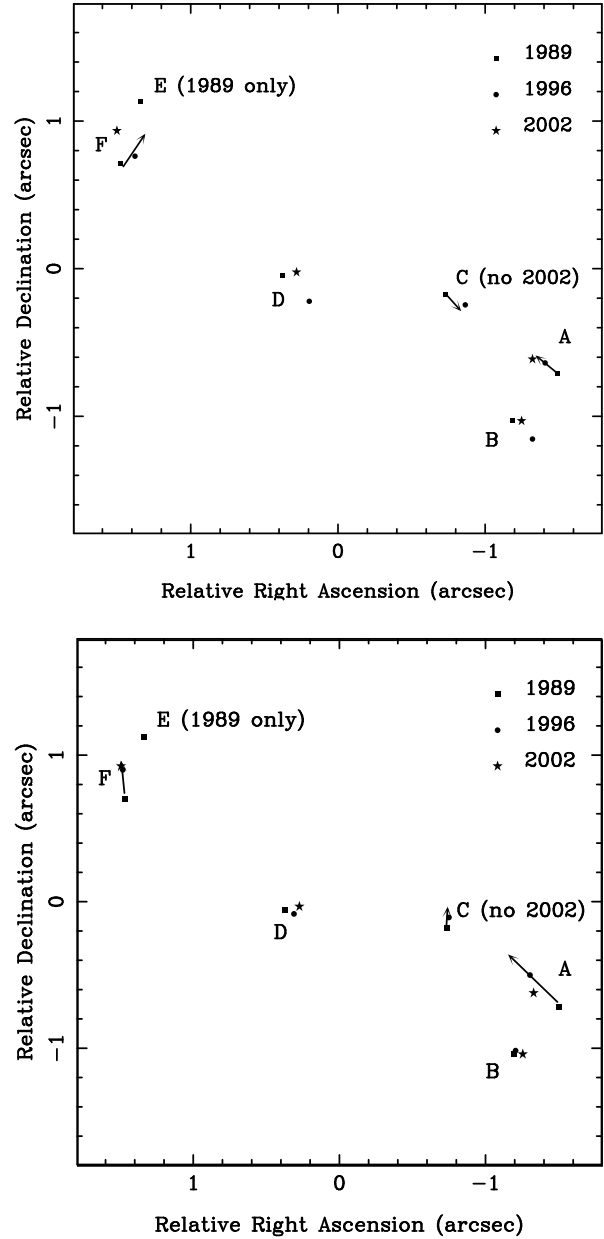

  \begin{center}
  
\includegraphics[width=8cm,angle=-90]{a1595f6a.ps}

\bigskip

\includegraphics[width=8cm,angle=-90]{a1595f6b.ps}

  \caption{Relative positions of radio sources in W3 IRS5
    before (top) and after (bottom) aligning sources B
    and D. Data are from \citet{claus94}, \citet{wilson03}, and
    Table~\ref{t:vla}. Arrows denote derived proper motions. 
    The symbol sizes represent the formal position error of 0\farcs06.} 
  \label{fig:motion}

  \end{center}

\end{figure}

\begin{table*}[tb]
  \caption{Proper motion solutions.}
  \label{tab:pm}
  \begin{center}
  \begin{tabular}{crrrr}

\hline
\hline
\noalign{\smallskip}
Source & \multicolumn{2}{c}{Offset (mas)$^a$} & \multicolumn{2}{c}{Proper motion (mas/yr)} \\
       & \multicolumn{1}{c}{$\alpha$} & \multicolumn{1}{c}{$\delta$} & \multicolumn{1}{c}{$\alpha$} & \multicolumn{1}{c}{$\delta$} \\
\noalign{\smallskip}
\hline

\multicolumn{5}{c}{\it Before Shift} \\ 
 A & --0.093 $\pm$ 11.550 &    0.158 $\pm$ 11.223          &   5.015 $\pm$ 1.801 &   8.202 $\pm$ 1.624 \\ 
 C & \multicolumn{1}{c}{...} & \multicolumn{1}{c}{...} & --7.813 $\pm$ 2.098 & --8.642 $\pm$ 2.112 \\ 
 F & --2.456 $\pm$ \phantom{0}8.277 & --27.419 $\pm$  7.630          & --5.359 $\pm$ 1.300 &  16.480 $\pm$ 0.806 \\ 
\multicolumn{5}{c}{\it After Shift} \\ 
 A &    0.485 $\pm$ 11.550 &  16.559 $\pm$ 11.223          &   12.015 $\pm$ 1.801 & 23.176 $\pm$ 1.624 \\ 
 C & \multicolumn{1}{c}{...} & \multicolumn{1}{c}{...} &  --0.654 $\pm$ 2.098 & 10.242 $\pm$ 2.112 \\ 
 F &    0.082 $\pm$ \phantom{0}8.277 &  26.035 $\pm$  7.630          &    1.042 $\pm$ 1.300 & 16.340 $\pm$ 0.806 \\ 

\noalign{\smallskip}
\hline

  \end{tabular}
  \end{center}

$^a$ Best fit offset from the measured 1989 position; unavailable for
C where only two epochs were measured

\end{table*}

Figure~\ref{fig:motion} shows the positions of radio sources A...F derived by
us and by \citet{claus94} and \citet{wilson03}.  Sources B and D appear
stationary when comparing the 1989 and 2002 data, but seem to have shifted by
$\approx$0\farcs2 in the 1996 data.  This shift may be a systematic phase
error in the 1996 data. 
The likely cause is an atmospheric `wedge', or $\sim$100~km-sized
parcel of dense air partially covering the interferometer and causing
a gradient in atmospheric opacity (M.~Reid, priv.\ comm.).
By aligning the positions of sources B and D in 1996 with those of 1989 and 2002,
we derive a shift of $\Delta\alpha$ = 55$\pm$7 and $\Delta\delta$ = 146$\pm$36 mas.
However, since the magnitude of the shift is uncertain, we derive
proper motion values both before and after applying the shift
(Table~\ref{tab:pm}).

Source C was only detected at two epochs, which is sufficient to
estimate the magnitude and the direction of its motion.  For sources A
and F, three epochs are available, which allows us to solve
additionally for the position at the first epoch, using the least
squares technique. For source A, good fits ($\chi^2$/dof $\sim$1) to
the $\alpha$ and $\delta$ motions are obtained if no shift is applied.
However, for source F, the fits are poor ($\chi^2$/dof $>$10) whether
the shift is applied or not. One possible explanation for these poor
fits are deviations from the assumed uniform motions.  The 1996 data
thus may confirm the proper motion of component F found by
\citet{wilson03} and show that components A and C may move as well,
but do not allow precise measurement of the magnitude and direction of
the motions.

If radio sources A, C and F are internally ionized, their ionizing stars
  must be moving along, since at an electron density of
  3$\times$10$^6$~\ccm, the recombination time scale is
  $\sim$1~month while the sources are seen over several years. 
The free-fall speed in the gravitational potential of the
  molecular gas ($M$=260~\msol, $R$=60,000~AU: \citealt{fvdt00}) is
  $\approx$3~\kms. The potential of the star cluster can be estimated
  by $M$=20~\msol\ and $R$=1000~AU (the typical separation of the
  radio sources: Fig.~\ref{f:vla}), which gives a free-fall speed of
  $\approx$6~\kms. The derived proper motions of the radio sources are much
  faster than these values, and therefore would not represent
  bound motions. Perhaps these stars were ejected from the
  cluster in a close stellar encounter, and are very young runaway OB
  stars, like the BN object in Orion, which is moving at 50~\kms\ \citep{plambeck95}.

We conclude that the evidence for proper motions remains weak,
even with three epochs measured. This shows graphically in
Fig.~\ref{fig:motion}: the 1996 positions do not generally lie between
those for 1989 and 2002. Quantitatively, it shows in the large error
margins in Table~\ref{tab:pm}. The next section explores alternative
explanations for the transient radio sources in W3~IRS5.

\subsection{Transient radio sources: Shocked clumps?}
\label{sec:shock}

The transient radio sources of W3~IRS5 may also be explained by shocks
which occur when the winds from the young O-type stars hit clumps in
the surrounding molecular material. Such a picture of massive star
formation has been described by, e.g., \citet{franco90} and \citet{dyson02}. 
Observational support has been found in the source Cep~A \citep{hughes01},
which is of somewhat lower luminosity and distance than W3~IRS5.

To estimate the radio emission from wind-shocked clumps, we use the
model by \citet{hollenbach89}.  In the limit that the stellar wind is
much less dense than the molecular clump, the flux $F_i$ of ionizing
photons is given by

$$F_i = n_0 v_S F(v_S)$$

where $n_0$ is the density of the clump, $v_S$ the shock velocity, and
$F(v_S)$ the fractional ionization of the shocked gas. To have
$F(v_S) \sim 1$, the shock must fully dissociate the \hh\ clump and heat it
to $\sim$10$^5$~K, so that it emits ionizing photons. 
The required shock velocity is $\sim$100~\kms; 
at lower velocities, $F(v_S)$ drops exponentially due to the
Boltzmann distribution. 
Velocities of $\sim$100~\kms\ are commonly observed for the winds of
deeply embedded high-mass stars including W3~IRS5, both in hydrogen
recombination lines \citep{bunn95} and in CO mid-infrared absorption
lines (\citealt{mitch91,fvdt99}).

The emission measure of the shock-ionized clump is

$$ n_e^2 l = \frac{F_i}{\alpha_B} \approx
   10^8 \left( \frac{n_0}{10^7 {\rm cm}^{-3}} \right)
        \left( \frac{v_S}{100 {\rm km\, s}^{-1}} \right) 
   {\rm cm}^{-6}{\rm pc} $$

where $\alpha_B$ is the Case~B recombination coefficient (\S~\ref{sec:ionize}).
The observed values of $l\ltsim$100~AU and $n_0$=10$^6$~\ccm\ agree
with this prediction within order of magnitude. We conclude that winds
from young O-type stars shocking and ionizing clumps in the ambient
cloud provide a viable model for the transient radio emission observed
in W3~IRS5.

\subsection{Diffuse emission: Envelope structure}
\label{sec:envelope}

Flux densities for the diffuse mid-infrared emission can be obtained by
subtracting the point source contributions (columns 4, 6 and 10 of
Table~\ref{t:ir}) from the total flux density (column 14). The emission is
roughly elliptical in shape, with the major axis more or less aligned with the
line connecting MIR1 and MIR2. Going from short to long wavelengths, the axis
ratio (measured at the 1\% level) decreases from $\approx$1.6 to $\approx$1.0,
while the radius (the average of the semi-major and semi-minor axes) increases
from $\approx$2300 to $\approx$4100~AU.

The brightness distribution of the diffuse emission is consistent with
heating by sources MIR1 \& MIR2. Short wavelengths probe warm dust
close to the individual stars, so that the emission has two peaks.
Longer wavelengths probe cooler dust that is far enough away that the
distances to both stars are about the same, leading to round contour
shapes. 

At a size of $\gtsim$1000~AU, the diffuse emission cannot be optically
thick: even for temperatures as unrealistically low as 100~K, the
far-infrared luminosity is exceeded, even with zero foreground
extinction.
The envelope must also have a low mid-infrared optical depth
to give us a view of the central objects.

The envelope of W3~IRS5 was modeled by \citet{campbell95}, based on
far-infrared data, and by \citet{fvdt00}, based on submillimeter data.
Assuming single power laws for the density structure,
these models have $\tau$(100~\mic)$\sim$1, or $\tau$(10~\mic)$\sim$40,
 much higher than estimated above.
The W3~IRS5 core is embedded in a large-scale molecular cloud, and the
observed submillimeter emission could contain contibutions from the
background cloud, but not more than 50\%. Neither the mid-infrared
nor the submillimeter images suggest deviations from spherical
symmetry stronger than a modest flattening (axis ratio $<$2). 

To reconcile the submillimeter and far-infrared data with the
mid-infrared data, one may explore broken power laws for the density
distribution, or a combination of dense shells and power laws.

\section{Conclusions and Outlook}
\label{concl}

\begin{figure*}[th]
  \begin{center}
\includegraphics[width=12cm,angle=-90]{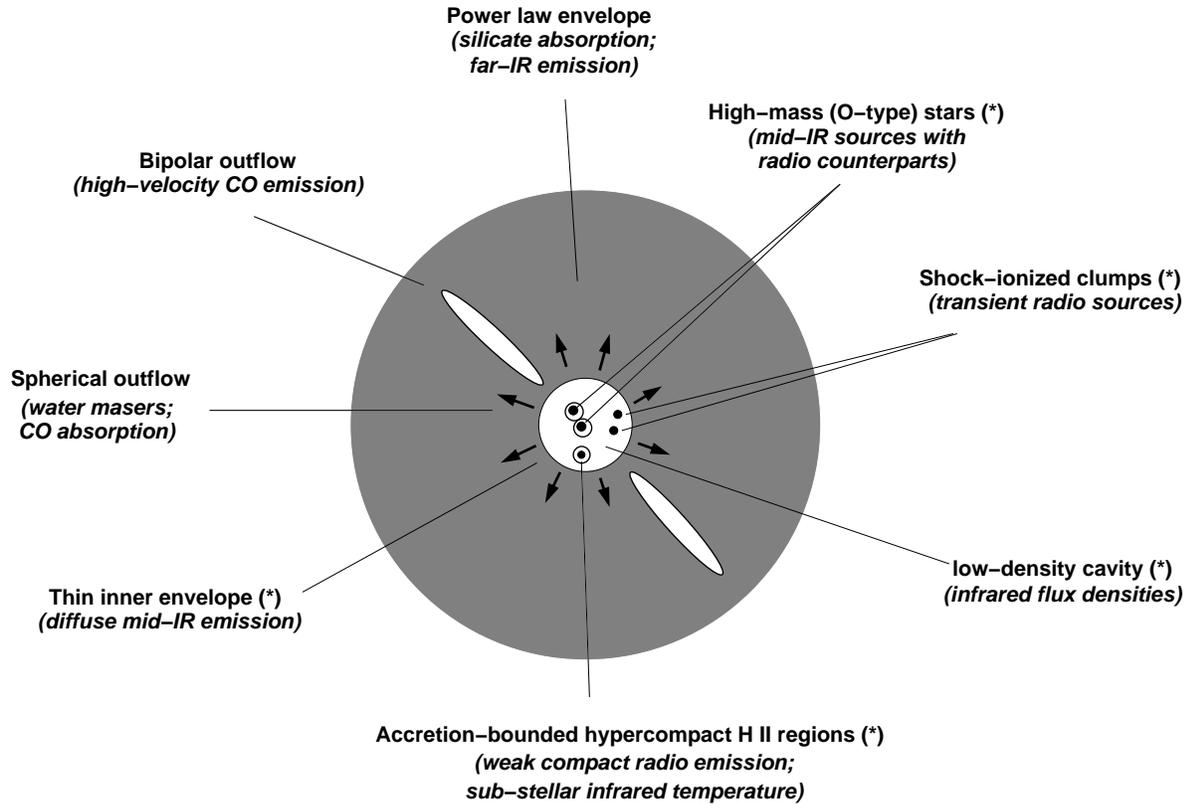}  

  \caption{Schematic view of the W3 IRS5 region, as projected on the
    sky, with the observational characteristics of each physical
    component indicated. Asterisks mark new findings of this paper.}

  \label{fig:cartoon}
  \end{center}
\end{figure*}

Observations at subarcsecond resolution at mid-infrared and radio
wavelengths have led to a detailed picture of the W3~IRS5 region,
shown schematically in Figure~\ref{fig:cartoon} and described below.

\begin{itemize}
\item The two bright mid-infrared sources with radio emission probably
  are deeply embedded high-mass stars.  They are both close to groups
  of \hho\ masers.  The measured mid-infrared diameters are
  consistent with blackbody emission at $T$=390~K and providing all of
  the far-infrared luminosity of W3~IRS5. 
\item  A third, weaker mid-infrared and radio source with associated
  \hho\ masers is probably a somewhat later-type star which is energetically
  unimportant. 
\item Radio source A has an optically thick radio spectrum, and may
  have a counterpart at long mid-infrared wavelengths
  ($\gtsim$17.65~\mic). It may be an extremely deeply embedded
  high-mass star.  The three power sources of W3~IRS5 then have
  $L$$\approx$40,000~\lsol\ each, which in the models of
  \citet{maeder89} makes them $\approx$20~\msol\ stars (ZAMS spectral
  type O8).

\item The region shows several transient radio sources.  Some
  of these may represent runaway OB stars, but most are probably
  clumps in the ambient material which are ionized and destroyed by
  shocks with the winds of the O-type stars. 

\item The low silicate optical depth suggests that no underlying
  silicate emission is present. This is most easily explained by a
  cavity separating the high-mass stars from their envelope. Perhaps
  the cavity was blown by the slow spherical outflow traced by the
  \hho\ masers.
\item The far-infrared and submillimeter emission, as well as the
  low-velocity CO mid-infrared absorption, arise in the large-scale
  envelope. The dense stellar cluster visible in the near-infrared
  (and X-ray) is embedded in the same envelope.
\end{itemize}

In the future, subarcsecond monitoring of W3 IRS5 at high radio
frequencies (VLA-A) is necessary to test the `proper motion' and
`shocked clump' hypotheses.  If proper motions are confirmed, the
shapes of the orbits will be a test of the `runaway star' hypothesis,
and will constrain the dynamics of this young cluster.  The emission
mechanism should be studied by simultaneous observations at three or
more wavelengths.  If the e-VLA does not provide the sensitivity
necessary to do this, ALMA will.

Mid-infrared imaging at wavelengths $\gtsim$20~\mic\ at subarcsecond
resolution is necessary to search for a mid-infrared counterpart to
radio source Q2=K3=A. The main requirements are higher sensitivity and
dynamic range than was achieved here; the higher angular resolution
offered by MIDI on the VLTI will be more useful to search for fine
structure.  Spatially resolved mid-infrared spectroscopy is necessary
to assign the high-velocity CO absorption features \citep{mitch91} to
particular stellar components. This may be a good project for
VLT/CRIRES. Future radiative transfer modeling efforts should consider
broken power laws for the density distribution in the envelope of
W3~IRS5.

\begin{acknowledgement}
  The authors thank David Hollenbach, Eric Keto, Ed Churchwell, Lee
  Hartmann, Tom Megeath, Mark Reid, 
  Enrik Kr\"ugel, Karl Menten, Tom Wilson, and
  Thomas Driebe for useful discussions. The staffs of the VLA
  (especially Claire Chandler) and Keck telescopes were helpful in
  assisting with the observations. We also thank Charles Townes, John
  Monnier, and Randy Campbell for help with the Keck observations.
\end{acknowledgement}

\bibliographystyle{aa}
\bibliography{w3irs5}

\end{document}